\documentclass[aps,prd,twocolumn,groupedaddress,showpacs,nofootinbib,final,superscriptaddress,floatfix]{revtex4-1}
\usepackage{graphicx}
\usepackage{dcolumn}
\usepackage{bm}
\usepackage{amsfonts}
\usepackage{amsmath}
\usepackage{amssymb}
\usepackage{subfigure}
\usepackage{color}
\newcommand{\red}[1]{{#1}}

\newcommand{\jcap}{JCAP}
\newcommand{\mnras}{MNRAS}

\newcommand{\h}{\ensuremath{\, {\rm h}}}

\newcommand{\Mpc}{\ensuremath{\,{\rm Mpc}}}
\newcommand{\K}{\ensuremath{\, {\rm K}}}

\newcommand{\Tr}{\ensuremath{\, {\rm Tr}}}

\begin{document}

\title{Precise measurements of inflationary features with 21 cm observations}

\author{Yidong Xu}
\affiliation{Key Laboratory for Computational Astrophysics, National Astronomical Observatories, Chinese Academy of Sciences,
 Beijing 100012, China}
\author{Jan Hamann}
\affiliation{Sydney Institute for Astronomy, The University of Sydney,
 Sydney NSW 2006, Australia}
\author{Xuelei Chen}
\affiliation{Key Laboratory for Computational Astrophysics, National Astronomical Observatories, Chinese Academy of Sciences,
Beijing 100012, China}
\affiliation{University of Chinese Academy of Sciences, Beijing 100049, China}
\affiliation{Center for High Energy Physics, Peking University,
 Beijing 100871, China}

\date{\today}

\begin{abstract}
Future observations of 21~cm emission using HI intensity mapping will enable us to probe the large scale structure of the Universe over very large survey volumes within a reasonable observation time.  We demonstrate that the three-dimensional information contained in such surveys will be an extremely powerful tool in searching for features that were imprinted in the primordial power spectrum and bispectrum during inflation.  Here we focus on the ``resonant'' and ``step'' inflation models, and
forecast the potential of upcoming 21~cm experiments to detect these inflationary
features in the observable power- and bispectrum.  We find that the full scale Tianlai experiment and the Square Kilometre Array (SKA) have the potential to improve on the sensitivity of current Cosmic Microwave Background (CMB) experiments by several orders of magnitude.

\end{abstract}



\maketitle

\section{Introduction}

While generic slow-roll models of cosmic inflation predict a nearly scale-invariant power spectrum of primordial curvature perturbations, there exist also many theoretically motivated implementations of the inflationary mechanism that predict \emph{features}, i.e., significant local deviations from scale invariance~\cite{Chluba:2015bqa}.  Power spectrum features are typically accompanied by a correlated, similarly strongly scale-dependent signal in higher-order spectra (e.g., \cite{2007JCAP...06..023C,2012JHEP...05..066A,2013JCAP...10..038B}),
which in principle allows us to discriminate between different scenarios by combining power spectrum and bispectrum information~
\cite{2015PhRvD..91b3502F,2016PhRvD..93d3536M}.  
However, analyses of present cosmic microwave background (CMB) anisotropy data have not found evidence for such features in the power spectrum~\cite{2015arXiv150202114P} or bispectrum~\cite{2015arXiv150201592P,2016PhRvD..93d3536M,2015PhRvD..91l3506F,2015arXiv151208977A} with a statistical significance higher than 3$\sigma$, after accounting for the look-elsewhere effect.  
It is therefore worth enquiring whether other observables may be more suitable for the detection of such features.

CMB lensing aside, the temperature and polarization maps of the CMB only provide us with 2-dimensional information about cosmic perturbations.  This not only imposes a fundamental limit to how precisely we can predict the angular power spectra (i.e., cosmic variance), but also obscures features through the necessary geometrical projection effect.  The large scale structure (LSS) of the Universe, on the other hand, is accessible to tomographic measurements, which retain the 3-dimensional information of the perturbations.  For a sufficiently large survey volume, cosmic variance can be pushed beyond the CMB limit.  Constraints on features models have previously been discussed in the context of using the galaxy power spectrum~\cite{2011PhRvD..84h3505C,2012JCAP...04..005H,2015PhRvD..91f4039H}.  

Here we investigate the potential of detecting inflationary features in primordial density perturbations using sky surveys with the 
redshifted 21 cm emission from neutral hydrogen (HI), especially the 21cm intensity mapping observations. In the 
intensity mapping mode of observation,  individual galaxies or clusters are not resolved,  only the total 21cm intensity of large cells 
which contains many galaxies are measured \cite{2008PhRvL.100i1303C}. What the intensity mapping survey loses in angular 
resolution it makes up for in survey speed, allowing us to potentially cover unprecedented survey volumes, and it has been shown to have 
exquisite sensitivity to various cosmological parameters \cite{2015ApJ...803...21B, 2015ApJ...798...40X}. 
A number of  21cm intensity mapping projects have been proposed, 
such as the single dish array feed BINGO (BAO from Integrated 
Neutral Gas Observations) project \cite{2013MNRAS.434.1239B}, and the cylinder arrays
CHIME (Canadian Hydrogen Mapping Experiment) \cite{CHIME} and Tianlai (Chinese for ``heavenly sound") 
projects  \cite{2012IJMPS..12..256C}.  Intensity mapping survey 
is also being considered for the upcoming Square Kilometer Array (SKA) phase one mid-frequency dish 
array (SKA1-MID) \citep{2015MNRAS.450.2251Y}.  Below we shall study the full scale Tianlai array and the SKA1-MID cases.  
We investigate two observables: the 21 cm \emph{power spectrum} and \emph{bispectrum} respectively, and focus on two models with oscillatory features: the \emph{resonant} model and the \emph{step} model.

\section{The Resonant and Step Models}

Representative for cases with features extending over the entire range of observable scales, 
we consider the resonant model \citep{2008JCAP...04..010C}. 
The resonant model may be realized in many different contexts including the axion monodromy scenario 
\cite{2010PhRvD..82d6003M}, where the inflaton field is modulated by a sinusoidal oscillation of frequency $\omega$.  
The power spectrum is given by \cite{2011JCAP...01..017F}
\begin{equation}\label{Eq.Pres}
P^{\rm res}_{\Phi}(k) = P_{\Phi}(k)\, \left[1\,+\, \frac{8\,f^{\rm res}}{C_\omega^2}\, 
\cos \left(C_\omega\,\ln \frac{k}{k_{\rm p}}\right) \right],
\end{equation}
where $f^{\rm res}$ describes the amplitude of the resonant non-Gaussianity,  $k_{\rm p}$ is the pivot scale 
which we fix to $k_{\rm p} = 0.02 \Mpc^{-1}$, and $C_\omega \equiv \omega/H_{\rm I}$ is the resonance ``frequency", 
$H_{\rm I}$ is the Hubble parameter during inflation. For axion monodromy inflation, 
the observed amplitude of the power spectrum imposes a limit of
$f^{\rm res}\lesssim 10^{-3} \, C_\omega^{5/2}$~\cite{2010JCAP...06..009F,2010PhRvD..82f3517M}. 
The corresponding bispectrum reads~\cite{2011JCAP...01..017F,2011PhRvD..84h3505C}
\begin{eqnarray}\label{Eq.Bres}
&&B^{\rm res}_{\Phi}(\vec{k}_1,\vec{k}_2,\vec{k}_3) = \frac{80\pi^4}{3}  
 \frac{f^{\rm res}\, \Delta_\Phi^2\, }{k_1^2 k_2^2 k_3^2}\, \times \left[\sin \left(C_\omega\, \ln \frac{K}{k_{\rm p}} \right) + \vphantom{\sum_{i\neq j}\frac{k_i}{k_j}\,
+ \, \mathcal{O}\left(\frac{1}{C_\omega^2} \right) } \right. \nonumber \\
&& \left.
\,+\,  \frac{1}{C_\omega}\, \cos (C_\omega\, \ln \frac{K}{k_{\rm p}} )\, \sum_{i\neq j}\frac{k_i}{k_j}\,
+ \, \mathcal{O}\left(\frac{1}{C_\omega^2} \right) \right].
\end{eqnarray}
where $k_i=|\vec{k}_i|$, $K = k_1+k_2+k_3$, and $\Delta_\Phi$ is the amplitude of primordial scalar power spectrum evaluated at $k_{\rm p}$.

Local features that affect only a relatively narrow $k$-range in the power spectrum and bispectrum 
can be generated, e.g., in models with brief rapid changes in the effective sound speed
 \cite{2011JCAP...01..030A,2012PhRvD..86l1301A,2014PhRvD..90b3511A}, or in models with a sudden 
step in the inflaton potential (step model) \cite{2001PhRvD..64l3514A,2007JCAP...06..023C,2008JCAP...04..010C}, 
and some other cases \cite{2006hep.th....7001A,2009JCAP...02..014A,2015EPJC...75..589C,2016EPJC...76..385C}.
In the latter case, the power spectrum can be approximated
analytically by~\citep{2012PhRvD..85b3531A,2013JCAP...10..038B}
\begin{equation}\label{Eq.Pstep}
\ln P^{\rm step}_{\Phi}(k) = \ln P_{\Phi}(k) \,-\, \frac{2}{3}\epsilon_{\rm step}\, 
W^{\prime}(k\,\tau_{\rm f})\, \mathcal{D}\left(\frac{k\,\tau_{\rm f}}{\beta}\right),
\end{equation}
where $W^{\prime} (x) \equiv \left(-\,3\,+\, \frac{9}{x^2}\right)\, \cos(2x) \,+\, \left(15 \,- \frac{9}{x^2}\right) \, \frac{\sin(2x)}{2x}$, 
and the damping function 
$\mathcal{D}(y) = \pi y/\sinh(\pi y)$ for a hyperbolic tangent step in the inflaton potential. 
The corresponding bispectrum is~\citep{2011PhRvD..84d3519A,2013JCAP...10..038B}
\begin{eqnarray}\label{Eq.Bstep}
&&B^{\rm step}_{\Phi}(k_1,k_2,k_3) = \frac{5}{12}\epsilon_{\rm step}\, 
\mathcal{D}\left(\frac{k\,\tau_{\rm f}}{2\,\beta}\right)\, \frac{(2\pi)^4\,\Delta_\Phi^2}{k_1^2\,k_2^2\,k_3^2}\, 
\nonumber \\
&&\times \left[\left(\frac{k_1^2+k_2^2+k_3^2}{k_1 k_2 k_3\, \tau_{\rm f}} \,-\, K\,\tau_{\rm f}\right)\, K\,\tau_{\rm f}\, \cos(K\,\tau_{\rm f}) \,-\,\right.
\nonumber \\
&& - \left.\left(\frac{k_1^2+k_2^2+k_3^2}{k_1 k_2 k_3\, \tau_{\rm f}} \,-\, 
 \frac{\sum_{i\neq j}k_i^2 k_j}{k_1 k_2 k_3}\, K\,\tau_{\rm f}\right) \, \sin(K\,\tau_{\rm f})\right].
\end{eqnarray}
Here $\epsilon_{\rm step} \ll 1$ is the height of the step in potential, 
$\beta \gg 1$ is the sharpness of the step, and $\tau_{\rm f}$ is the conformal time at which the step occurs. 
Larger values of $\beta$ imply a sharper step and 
thus a more extended shape of the feature envelope, making the signal easier to detect.

In either case, such features could be searched by measuring the power spectrum or bispectrum over a range of $k$. 
On large scales, the HI intensity traces the total matter density. As is usually done in such forecast, here we assume that the 
foreground can be removed, so that the measurement error on the 21 cm signal is determined simply by 
the system temperature, integration time, and the array configuration of the radio telescope. 
In redshift space, the power spectrum is smeared by the peculiar velocity, which we model as 
$P_{\rm s}(k; z) = [b_1^{\rm HI}(z) + f(z) \mu^2]^2\, e^{-k^2 \mu^2 \sigma_r^2} P_m(k; z)$,
where $b_1^{\rm HI}$ is the bias factor of HI, $f(z)$ is the linear growth rate, $\mu \equiv k_\parallel/k$ is the cosine of 
angle with respect to the line of sight, and 
$P_m(k; z)$ is the matter power spectrum at redshift $z$.
The non-linear dispersion scale, characterizing the ``Finger of God'' effect 
on small scales, is taken as $\sigma_r = 7\Mpc$ \citep{2007MNRAS.376..984L,2015ApJ...803...21B}.

\section{Forecasts}

We use the Fisher information matrix to forecast the expected measurement uncertainties. 
We take the 21cm power spectrum and bispectrum as our observables, and forecast the error in the measurement of  amplitude 
parameters for the feature models, such as $f^{\rm res}$ and $\epsilon_{\rm step}$, 
while keeping other parameters of the feature model, e.g. $C_\omega, \beta$ or $\tau_{\rm f}$ fixed in the forecast.  The likelihood is Gaussian,  
\begin{equation}
\mathcal{L}= [(2\pi)^n \det C]^{-1/2} \exp \left(-\frac{1}{2} \Delta C^{-1} \Delta\right),
\end{equation}
where $\Delta$ is the difference between data and prediction, $n$ is the number of data,  
and $C$ is the covariance matrix. Note that if we take $f^{\rm res}=0$ or $\epsilon_{\rm step}=0$ as the null hypothesis, 
the likelihood ratio used by a testing of the hypothesis of the presence of features in the data is given exactly by the same expression,
so the parameter forecast is equivalent to hypothesis testing. 
We shall also take the remaining cosmological parameters as fixed since they are uncorrelated with the 
feature parameters, and adopt the Planck-2015 model \cite{2015arXiv150201589P} as our fiducial cosmology model.
The Fisher matrix of the set of parameters of interest is then given by
$F_{\alpha\beta}=\frac{1}{2} \Tr[C_{,\alpha}  C^{-1} C_{,\beta}  C^{-1} ].$

In the forecast with power spectrum data, we found only a negligible difference when considering non-linear corrections. \red{For the bispectrum, the non-linear corrections is already
comparable to the amplitude of the primordial $f_{\rm NL}^{\rm eq.} \sim 1$ term on relatively large scales \cite{2016arXiv160200674B}. However,  only the mode-coupling part of the non-linear corrections should
be expected to have an impact on our ability to detect features, 
the non-mode-coupling corrections merely generate broad distortions but not oscillating features. The non-mode-coupling 
contribution is important if one is looking for physical effects that also predict a broad distortion, 
such as a non-zero neutrino mass, but much less relevant when it comes to looking for oscillatory 
features as we do in this paper. Neglecting them would bias the mean of the oscillation, but not affect much on the signal strength, so they do not greatly change the result of forecast. 
We will investigate the non-linear correction on the bispectrum in a subsequent study, here we use the tree-level bispectrum for the forecast.   }

The Fisher matrix $F^{\rm obs}_{\alpha\beta}$ for the power
spectrum and bispectrum is given in Ref.\cite{2015ApJ...798...40X}; here we reproduce the bispectrum case.
In terms of the reduced bispectrum $Q$, defined by 
$B(\vec{k}_1,\vec{k}_2,\vec{k}_3)\equiv Q(\vec{k}_1,\vec{k}_2,\vec{k}_3) [P(k_1) P(k_2) + {\rm perm.}],$
the Fisher matrix is
\begin{equation}
F^{\rm obs}_{\alpha \beta} \;\equiv\; \sum_{k_1=k_{\rm min}}^{k_{\rm max}}\, \sum_{k_2=k_{\rm min}}^{k_1}\, \sum_{k_3=k_{\rm min}^\star}^{k_2}\, \frac{\partial Q_{\rm s}}{\partial \alpha}\, \frac{\partial Q_{\rm s}}{\partial \beta}\, \frac{1}{\Delta Q_{\rm s}^2},
\end{equation}
where the three sums are over all combinations of $\vec{k}_1$, $\vec{k}_2$ and $\vec{k}_3$ that form triangles,
with $k_{\rm min}^\star = \max (k_{\rm min}, |k_1-k_2|)$.
The variance $\Delta Q_{\rm s}^2$  is approximately \cite{2007PhRvD..76h3004S}
\begin{equation}
\Delta Q_{\rm s}^2 (k_1,k_2,k_3) \;\simeq\; \frac{\Delta B_{\rm s}^2(k_1,k_2,k_3)}{\left[P_{\rm s}(k_1)P_{\rm s}(k_2) \,+\, (\rm perm.)\right]^2},
\end{equation}
where $\Delta B_{\rm s}^2$ is given by \citep{1998ApJ...496..586S}
$$\Delta B_{\rm s}^2(k_1,k_2,k_3) \simeq (2\pi)^3\, V_{\rm f}\, \frac{s_{\rm 123}}{V_{\rm B}}\, P_{\rm tot}(k_1)\, P_{\rm tot}(k_2)\, P_{\rm tot}(k_3).$$
Here, $P_{\rm tot}(k)=P_{\rm s}(k) + P_{\rm n}(k)$, $P_{\rm s}(k)$ and $P_{\rm n}(k)$ are the signal and 
noise power spectrum respectively, and  $V_{\rm f} \equiv k_{\rm f}^3 = (2\pi)^3/V$ is the $k$-space volume of the observation cells,
$V_{\rm B} \simeq 8\pi^2\, k_1\,k_2\,k_3\, \Delta k_1\, \Delta k_2\, \Delta k_3$ with $\Delta k_i = k_{\rm f}$, and 
$s_{\rm 123} = 6, 2, 1$  for equilateral, isosceles and general triangles, respectively. 

The range of oscillatory ``frequencies" that can be resolved by the power spectrum features is limited by the Nyquist-Shannon sampling theorem.  
On large scales (small $k$), the cutoff $k_{\rm min}$ and the resolution of $k$ measurement 
is determined by the volume of the respective redshift bin. Foreground removal may also 
reduce radial modes on larger scales.
On small scales (large $k$), the range of $k$ covered by the 21cm intensity mapping data is limited either by 
the angular resolution of the telescope $\theta_{\rm res}$ (with a Nyquist frequency given by $k_{\rm Ny} = \pi / \theta_{\rm res}$), 
or by a non-linear wavenumber cutoff, $k_{\rm nl}$, which we conservatively define by $\sigma(k_{\rm nl}, z) = 0.5$ in each redshift bin \citep{2003ApJ...598..720S}.  The radio experiments generally have sufficient frequency spectral resolution, so the radial 
direction is usually not a limiting factor, except for the smearing effect of the peculiar velocity, which is automatically taken into account by using the redshift space power spectrum.

 The full scale Tianlai array will be a 120 m $\times$ 120 m cylinder reflector array, covering the frequency range of 400-1420 MHz. 
 We assume a system temperature of 50 K and a sky area of about 10000 square degrees, with a total integration time of 1 year.  We divide 
the full frequency range into 8 bins of equal width. The corresponding noise power spectrum was given in Ref.~\cite{2015ApJ...798...40X}.  
The wavenumber range varies from about $0.025 - 0.11 \h\Mpc^{-1}$ at low redshifts,
to $0.001 - 0.16 \h\Mpc^{-1}$ at high redshifts.

The SKA1-mid includes a total of $N_d = 197$ dishes.  For simplicity we assume all of these to be 15 m dishes, 
though in reality 64 of them are 13.5~m MeerKAT dishes. 
The full frequency range of 350 -- 1420 MHz is divided into 9 bins in our calculation. 
As the SKA1-MID array is designed for multiple purposes, the short baselines are relatively few. To make intensity mapping 
observations, it has been proposed that the array is to be used as a collection of single dishes for observation on large scales 
by using the auto-correlation of each antenna, while the 
interferometry observation (cross-correlation between different antennas) is carried out concurrently to 
calibrate the receiver gain as well as observing at higher angular scales\cite{2015ApJ...803...21B}.  The noise power spectrum 
of the single dish (auto-correlation) data and the interferometer (cross-correlation) can be written as
\begin{eqnarray}
P_{\rm n}^{\rm auto} &=& \frac{T_{\rm sys}^2}{N_{\rm d}\, t_{\rm tot}}\, \Omega_{\rm survey}\, d_{\rm A}^2(z)\, y(z),\label{Eq.ska_noise_sd}\\
P_{\rm n}^{\rm cross} &=& \frac{T_{\rm sys}^2}{t_{\rm tot}\, n({\bf u})}\, \Omega_{\rm survey}\, \label{Eq.ska_noise_if}
\Omega_{\rm FoV}\, d_{\rm A}^2(z)\, y(z),
\end{eqnarray}
where $T_{\rm sys}=25\K$ is the system temperature, $t_{\rm tot}=10000$ hours is the total observation time we assumed,
$d_{\rm A}$ is the comoving angular diameter distance, $y(z)$ converts 
the observed frequency range into the radial distance, $\Omega_{\rm FoV}$ is the instant field of view of the dish,  
$\Omega_{\rm survey}=3\pi$ is the solid angle of the survey area we assumed,  and $n({\bf u})$ is the baseline 
number density on the {\it uv}-plane computed from the SKA1-MID array configuration \cite{ska-simple}.
The angular resolution of the single-dish limits the maximum $k$-range that this mode can  probe. The interferometer 
observation is limited to small scales, with $k_{\rm min}$ limited by the primary beam field of view, 
and $k_{\rm max}$ limited by the non-linear scale cutoff ($k_{\rm Nyq} \ll k_{\rm nonl}$). 
The total probed $k$ range is $0.0005 - 0.39 \h\Mpc^{-1}$.

\begin{figure}[t]
\centering{
\includegraphics[scale=0.35]{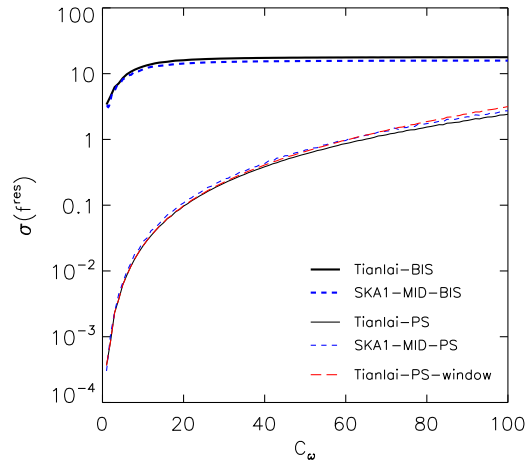}
\caption{The marginalized $1-\sigma$ error on $f^{\rm res}$ as a function of $C_\omega$
 in the resonant model for HI power spectrum
measurements {\it (thin lines)} and for HI bispectrum measurements {\it(thick lines)}, 
with the fiducial value of $f^{\rm res}$ set to 0.
In each set of lines, the solid and dashed lines are for Tianlai and SKA1-MID respectively. 
The thin long-dashed line shows the HI power spectrum measurement with Tianlai 
when the window function effect is taken into account.}
\label{Fig.resonance_sigfres}
}
\end{figure}

\section{Results}

The forecasted uncertainties on the amplitude of resonant non-Gaussianity, $f^{\rm res}$, 
are plotted in Fig.~\ref{Fig.resonance_sigfres} as a function of the resonance frequency.
The results shown here are for a fiducial value of $f^{\rm res} = 0$; 
we also tested different fiducial values and found that the choice of fiducial $f^{\rm res}$ 
only affects the result weakly.
The 1-$\sigma$ sensitivities to $f^{\rm res}$ derived from the HI power spectrum data are shown as thin lines,
while those from the HI bispectrum data are shown as thick lines.
In each set of lines, the solid and short-dashed lines are for Tianlai  and SKA1-MID respectively.
The difference between the full scale
Tianlai case and SKA1-MID is not large: both can make excellent measurement on the relevant redshift range and scales.

We find that $\sigma_{f^{\rm res}}$ increases with $C_\omega$, indicating that the test will be more sensitive to 
``low frequency" modulations.  The dependence on $C_\omega$ can be understood by looking at the actual amplitude of the 
modulations in the power and bispectrum: for the power spectrum, it is proportional to $f^{\rm res}/C_\omega^2$, so one expects $\sigma_{f^{\rm res}} \propto C_\omega^2$.  The bispectrum (Eq.~(\ref{Eq.Bres})) is dominated by the cosine term at low frequencies ($C_\omega \ll 10$).  Its amplitude scales with $C_\omega^{-1}$, yielding $\sigma_{f^{\rm res}} \propto C_\omega$, up to the point where the sine term of Eq.~(\ref{Eq.Bres}), which is independent of $C_\omega$, takes over, and the sensitivity approaches a constant value.
Within the range of $C_\omega$ considered by us, the HI power spectrum observations always have better sensitivity to the amplitude of resonant non-Gaussianity than the bispectrum observations.  At very high ``frequencies" ($C_\omega \gg 100$), the more favourable scaling of the bispectrum's sensitivity may invert the situation, though there the $k$-space resolution limit applies.
The bispectrum measurements could achieve 
$\sigma_{f^{\rm res}}\lesssim 18$ for Tianlai and $\sigma_{f^{\rm res}}\lesssim 16$ for the SKA1-MID,
and the power spectrum measurements could achieve (for $C_\omega \lesssim 100$)
$\sigma_{f^{\rm res}}\lesssim 2.5$ for Tianlai and $\sigma_{f^{\rm res}}\lesssim 2.8$ for the SKA1-MID.

 \citet{2015PhRvD..91d3534M} predicted the 1-$\sigma$ error on $f^{\rm res}$ from CMB bispectrum measurement 
 to be  $\sim 300 - 3000$ for $C_\omega \lesssim 100$ (cf.\ Fig.~8 in Ref.~\cite{2015PhRvD..91d3534M}).
We note that even with the bispectrum measurement from 21~cm intensity mapping, the constraints on
$f^{\rm res}$ in the resonant model can be more than two orders of magnitude better than
those of the CMB, and even stronger constraints can be
obtained from the HI power spectrum data, particularly for small $C_\omega$. 

\begin{figure*}[t]
\centering{
\includegraphics[scale=0.35]{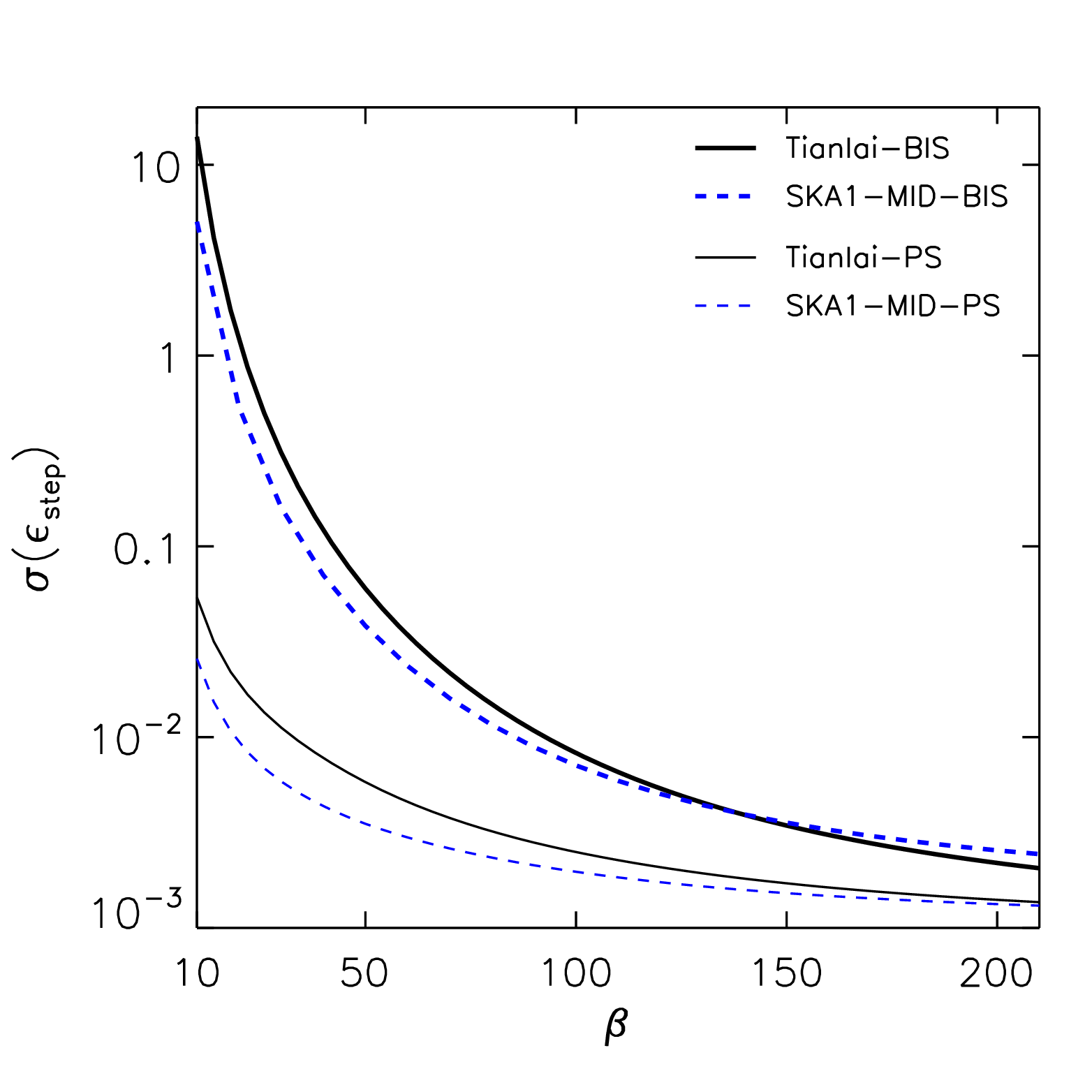}
\includegraphics[scale=0.35]{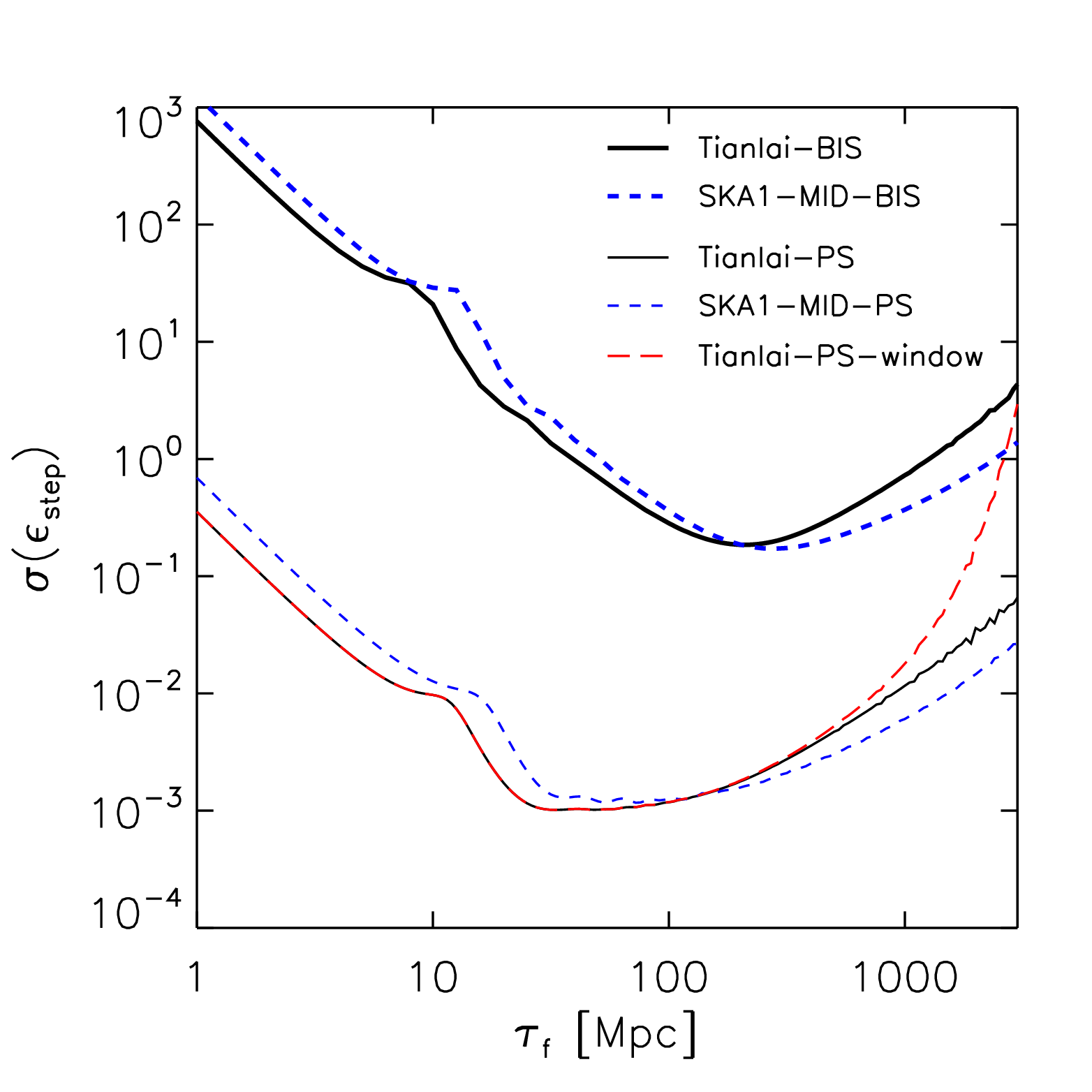}
\caption{The marginalized $1-\sigma$ error on $\epsilon_{\rm step}$ in the step model for HI power spectrum
measurement {\it (thin lines)} and for HI bispectrum measurements {\it(thick lines)}. 
In each set of lines, the solid and dashed lines are for Tianlai and SKA1-MID respectively. 
{\it Left panel:} The predicted $\sigma_{\epsilon_{\rm step}}$ as a function of $\beta$ for $\tau_{\rm f} = 1440 \Mpc$. 
{\it Right panel:} The predicted $\sigma_{\epsilon_{\rm step}}$ as a function of $\tau_{\rm f}$ for $\beta = 20$. 
The thin long-dashed line shows the HI power spectrum measurement with Tianlai 
when the window function effect is taken into account.
The fiducial value of $\epsilon_{\rm step}$ is set to 0.}
\label{Fig.step_sigestep}
}
\end{figure*}

The constraint on the height of the step in the inflaton potential is plotted in Fig.~\ref{Fig.step_sigestep}. 
The left panel shows $\sigma_{\epsilon_{\rm step}}$ as a function of sharpness $\beta$, 
for a given step position $\tau_{\rm f} = 1440 \Mpc$.  For $\beta \gtrsim 10$, 
the HI bispectrum measurements could achieve 
$\sigma_{\epsilon_{\rm step}}\lesssim 14$ for Tianlai, and
$\sigma_{\epsilon_{\rm step}}\lesssim 5.0$ for SKA1-MID;
while the HI power spectrum measurements could achieve 
$\sigma_{\epsilon_{\rm step}}\lesssim 0.054$ for Tianlai, and
$\sigma_{\epsilon_{\rm step}}\lesssim 0.026$ for  SKA1-MID.
Since sharper features are accompanied by a more extended envelope, the sensitivity increases with larger $\beta$.
However, we note that the theory is strongly coupled for $\beta > 170$ \citep{2014PhRvD..89h3531A}.
The right panel shows the $\sigma_{\epsilon_{\rm step}}$  as a function of $\tau_{\rm f}$, for $\beta = 20$. 
Since $\tau_{\rm f}$ determines the position of the feature in the spectra, the shape of the curves simply reflects 
the fact that the data will be most sensitive around $k \gtrsim 0.1 \Mpc^{-1}$ for the power spectrum and 
around $k \gtrsim 0.05 \Mpc^{-1}$ for the bispectrum. 
In the step model the HI power spectrum measurement will be sensitive to sub-percent modulations, and for sufficiently sharp steps, this is also 
true for the bispectrum.  Similar to the resonant model, the sensitivity of the bispectrum data to $\epsilon_{\rm step}$ is somewhat lower than that 
of the power spectrum data. For a step feature with $\tau_{\rm f} \approx 1440 \Mpc$ the SKA1-MID would have a slight edge in sensitivity over Tianlai.

\subsection{The effect of window function}
The measurement of power spectrum is affected by the window function, which depends on the survey volume.
In Fig.~\ref{Fig.resonance_sigfres} and the right panel of Fig.~\ref{Fig.step_sigestep}, we also show the predictions
 (plotted with a thin long-dashed line in each plot) for the HI power spectrum measurement with Tianlai 
when the $k$-space window function effect is taken into account. 
For the resonance model, it turns out that over the range of resonance frequencies considered here, 
the window function effect is not important for Tianlai, 
and completely negligible for SKA1-MID (not shown in the figure).
This is because the window function operates in $k$, not in $\log k$.  So for a logarithmic oscillation 
in the resonance model, the effect of the window function will be strongly scale-dependent. 
For the values of $C_\omega$ we considered, the ``surviving'' part of the oscillations is always enough to 
dominate the signal, and there is no significant loss of sensitivity.
The survey volume is large enough to guarantee a sufficiently high resolution 
in $k$-space for primordial resonant features not to get smeared out in the observed spectra. 
For the step model, the situation is different because the oscillations have a constant frequency in $k$-space. 
If the frequency is high enough (i.e., if $\tau_f$ is large enough), the signal is going to be smeared out 
on the entire range of observable scales.
The figure shows that this happens around $\tau_f \gtrsim 1000\Mpc$.
The limited survey volume could severely reduce the measurement precision for step models
with larger $\tau_f$.

\subsection{The effect of foregrounds}
\begin{figure*}[t]
\centering{
\includegraphics[scale=0.28]{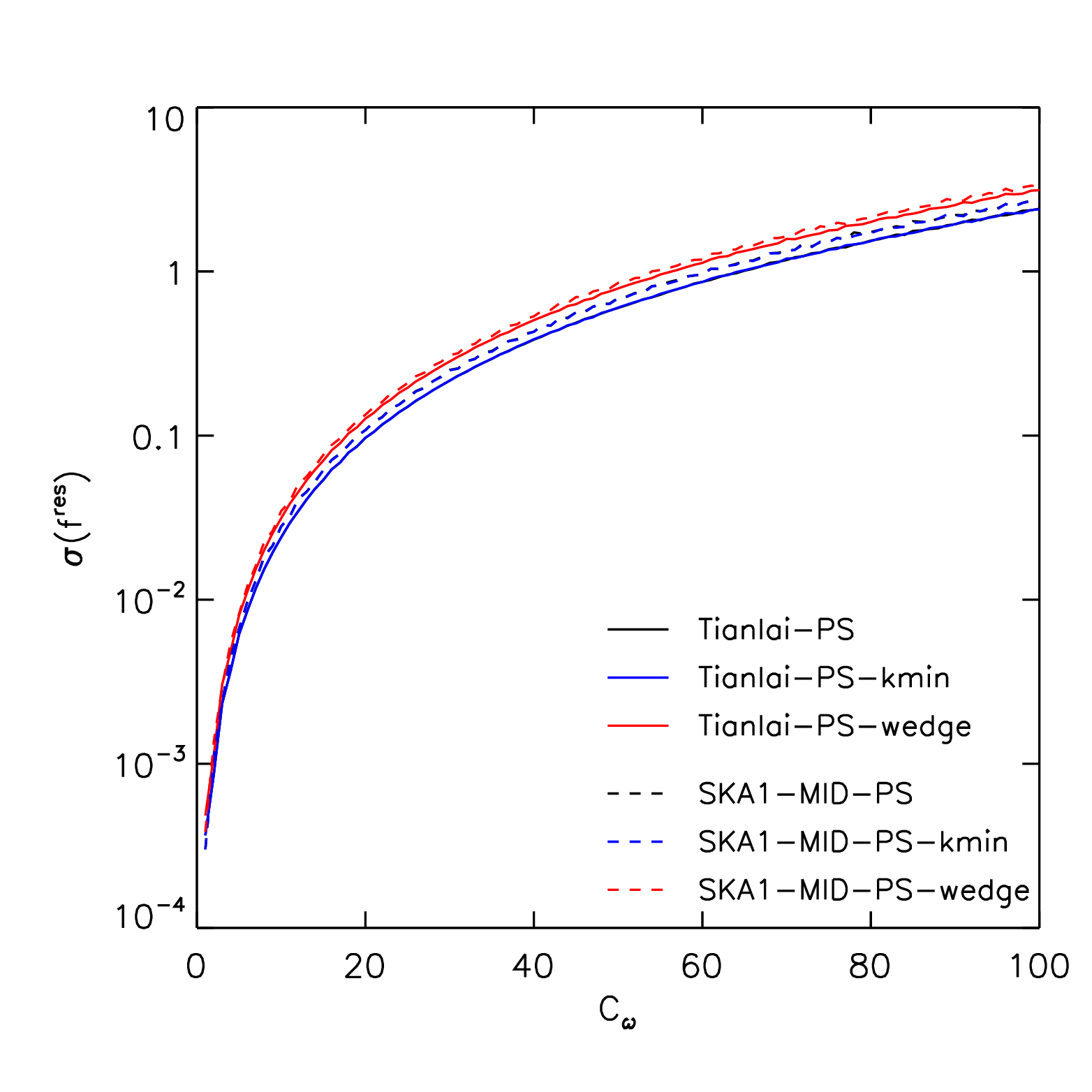}
\includegraphics[scale=0.28]{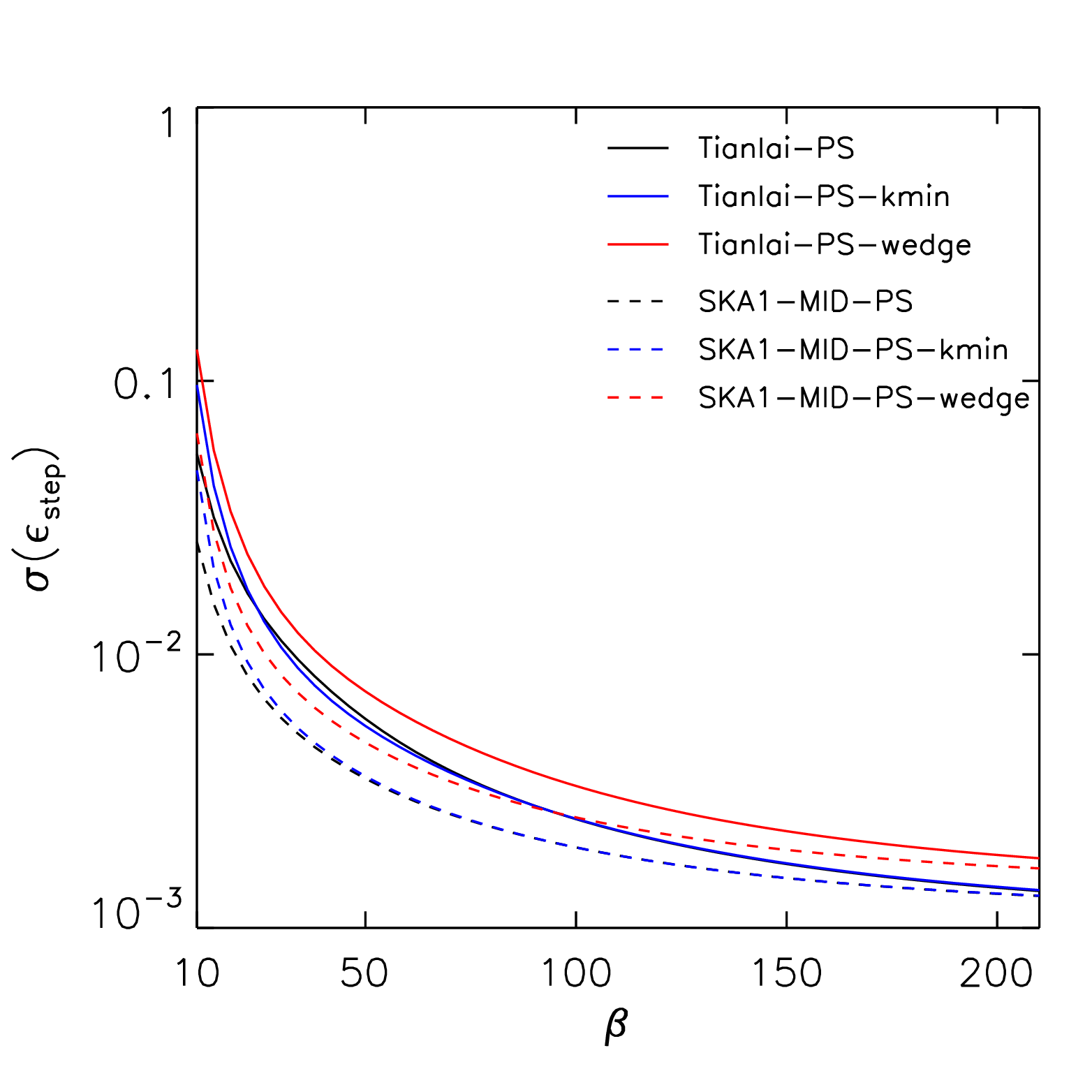}
\includegraphics[scale=0.28]{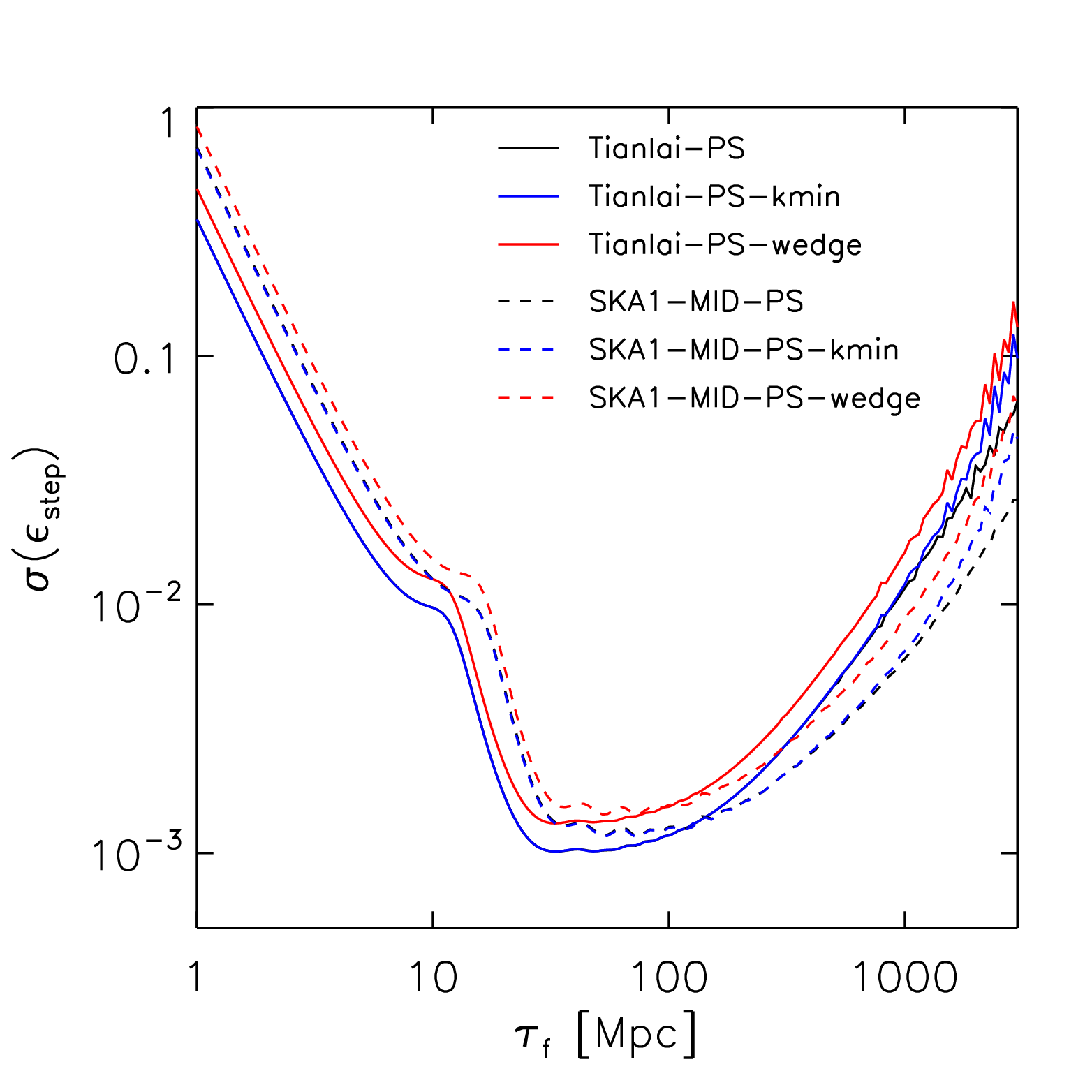}
\caption{The effect of foreground contamination for HI power spectrum measurements.
{\it Left panel:} The marginalized $1-\sigma$ error on $f^{\rm res}$ as a function of $C_\omega$
 in the resonant model.
{\it Central panel:} The marginalized $1-\sigma$ error on $\epsilon_{\rm step}$ as a function of 
$\beta$ for $\tau_{\rm f} = 1440 \Mpc$ in the step model.
{\it Right panel:} The marginalized $1-\sigma$ error on $\epsilon_{\rm step}$ as a function of 
$\tau_{\rm f}$ for $\beta = 20$ in the step model.
In each panel, the solid and dashed lines are for Tianlai and SKA1-MID respectively. 
In each set of lines, the black, blue, and red lines (from bottom to top) 
correspond to the cases with no foreground contamination,
with a $k_{\rm min}$ cut-off at $0.01\h\Mpc^{-1}$, and with both the $k_{\rm min}$ cut-off and the wedge exclusion, 
respectively.}
\label{Fig.foreground_effect}
}
\end{figure*}

Real observations of the large scale structure with the 21 cm intensity mapping are very 
challenging due to the bright Galactic and extra-galactic foregrounds, though various foreground 
removal and calibration techniques are being developed. In the above we assume that the foregrounds can be removed perfectly.
However, the foreground removing procedures which make use of the spectral smoothness of the foreground radiation would generally 
unable to recover some Fourier modes with small radial wave numbers \cite{2008PhRvL.100i1303C}, and 
contamination from the chromatic instrument response would result in a ``foreground wedge'' in $k$-space
\cite{2010ApJ...724..526D,2012ApJ...745..176V,2012ApJ...752..137M}. 
Here we investigate the effect of foregrounds contamination in an approximated way.

For cylinder array such as Tianlai and CHIME,  only modes with $k \gtrsim 0.01$ $(0.03) \h\Mpc^{-1}$ could be 
used at $z=1.2$ ($z=2$) \cite{2008PhRvL.100i1303C}. To test the effect of losing the small wavenumber modes, 
we calculate the constraints with a simple cutoff at $k_{\rm min} = 0.01 \h \Mpc^{-1}$ for the whole redshift range 
probed as a pessimistic estimate. The results for the HI power spectrum measurements are plotted with blue lines in Fig.~\ref{Fig.foreground_effect}, and the fiducial constraints without foreground contamination are plotted
with black lines for comparison. We find that for the resonance model, the blue lines overlap the black lines, indicating that losing the small 
wavenumber modes has almost no impact on the measurement precision. As for the step model,
on the other hand, losing the small wavenumber modes does affect the small $\beta$ and large $\tau_f$ ends,
especially increasing the measurement error for $\tau_f \gtrsim 2000 \Mpc$.

The effect of the ``foreground wedge'' can be modeled roughly as losing a fraction of $\mu_{\rm min}$ of the Fourier 
modes in the Fisher forecast formalism \cite{2016MNRAS.456.3142S}. Now $\mu_{\rm min}$ is determined by the edge of the wedge, i.e.
\begin{equation}
\mu_{\rm min} = \frac{k_\parallel}{\sqrt{k_\perp^2 + k_\parallel^2}}
= \frac{d_{\rm A}(z) H(z)/ [c(1+z)]}{\sqrt{1 + \{d_{\rm A}(z) H(z)/ [c(1+z)]\}^2}}.
\end{equation}
Here $k_\parallel$ and $k_\perp$ are the line-of-sight and transverse wavenumber respectively, and
$H(z)$ is the Hubble parameter.
To test the effect of losing information in the foreground wedge, we further retain a fraction of $(1 - \mu_{\rm min})$
of the Fourier modes in the Fisher matrix, and plot the resultant constraints with the red lines in
Fig.~\ref{Fig.foreground_effect}.
We find that the effect of the ``foreground wedge'' is obvious but not significant for both the resonant and step 
models, so we conclude that even in the presence of foreground contamination, the 21 cm intensity mapping 
observations of the LSS with Tianlai and SKA1-MID could still put tight constraints on the feature models.

\red{Finally, the non-linear corrections may limit the usable modes to small $k$. We tested this effect
by computing the limits with half value of $k_{\rm max}$. For the resonance model, the constraints at 
different $C_\omega$ are almost equally affected, and the $\sigma(f^{\rm res})$ values 
are increased by less than a factor of two. For the step model, at small $\tau_f$ the sensitivity derived from the bispectrum is reduced by a factor of a few, as the information at small scales are lost. At larger $\tau_f$ the 
sensitivity is almost not affected. Similarly, a larger $\beta$ will lead to a more extended feature with a wider envelope, therefore losing the largest $k$-modes will lead to greater loss of sensitivity for larger $\beta$, again up
to factor of a few.
}

\section{Conclusion}

Very recently the potential of greatly improving constraints on oscillatory features in power spectrum with future large scale 
structure observations was noted in Refs.~\cite{2016arXiv160509365C,2016arXiv160603747B}, which investigated the potential of 
Euclid and LSST galaxy power spectrum observations, and Ref.~\cite{2016arXiv160509364C}, which looked at future 21~cm 
measurement through the dark ages. 
Here we show that the upcoming 21 cm intensity mapping observations of the LSS in the post-reionization Universe {\it alone} could put extremely tight constraints on the feature models.
While the exact limit derived from the observation may depend on the details of the survey, such as the 
redshift range, sky area, system temperature and total observation time, and the precision actually achieved may be somewhat 
lower than the forecast due to less-than-perfect foreground removal, these surveys would still make orders-of-magnitude improvements 
over the two-dimensional CMB measurements. Furthermore, we also considered the bispectrum measurements, which were not previously
considered for galaxy surveys, and found that it could also provide constraints better than the CMB.
In addition, the sensitivity may be further improved by combining the power spectrum and bispectrum measurements. \\\\

\section*{Acknowledgements}
We thank Gary Shiu, Peter Adshead, and Xingang Chen for helpful discussions.
YX is supported by the NSFC grant 11303034, and the Young Researcher Grant of
National Astronomical Observatories, Chinese Academy of Sciences.
JH gratefully acknowledges the support of a Future Fellowship of the Australian Research Council, 
and wishes to thank the Kavli Institute for Theoretical Physics China and the National Astronomical Observatory of China for their hospitality.
 XC is supported by the MoST 863 program grant 2012AA121701, the CAS strategic Priority Research Program XDB09020301, the CAS grant
 QYZDJ-SSW-SLH017, and NSFC grant 11373030.

\bibliography{references}

\end{document}